# THE PROFESSIONALIZATION OF THE HACKER INDUSTRY

Tyson Brooks

Syracuse University, USA

*ABSTRACT*

*Society is inextricably dependent on the Internet and other globally interconnected infrastructures used in the provisioning of information services. The growth of information technology (IT) and information systems (IS) over the past decades has created an unprecedented demand for access to information. The implication of wireless mobility are great, and the commercial possibilities of new and innovative wireless flexibility are just beginning to be realized through the emergence of the Internet of Things (IoT). This article takes a look the history of hacking and professionalization of the hacker industry. As the hacker industry becomes more fully professionalized, it is becoming much more adaptive and flexible, making it harder for intelligence and law enforcement to confront. Furthermore, the hacker industry is blurring the distinction between motivated crime and traditional computer security threats - including the disruption of critical infrastructures or the penetration of networks.*

## 1. HACKING

What is the most important computing-related phenomenon of the last couple of decades? Contrary to the headlines, the most significant emergent pattern may be neither the collapse of the dotcom bubble, nor the emerging hegemony of Google since then, nor even the rise of Web. 2.0 and social networking applications, from Facebook [i.e., Meta] to Twitter. Rather, the most consequential computing phenomena of the last decade may be the Internet of Things (IoT) and the evolution of a professional hacker "industry" that is both global in scope and diversified at a local and regional level. Malicious software (malware), ransomware, and cybercrime, which are the products of this industry, not only threaten United States national security in direct and indirect ways, but also risk compromising the integrity of the Internet as a commercial and communications platform. While the threat posed by hacking has received attention from both the media and the United States government, it has to date been discussed largely in technical terms, both by government officials and security vendors. As part of a larger effort by security professionals to understand the global cultures of hackers, this article seeks to describe the rise and diversification of the hacker.

Recent years saw some of the largest, most sophisticated, and most severe cyber-attacks, such as WannaCry [48], the Equifax data breach [35], and the Facebook data leak [44], which affected millions of consumers, government entities, and thousands of businesses. For American audiences, the term "hacking" is loaded with a variety of connotations, ranging from positive ones associated with the open-source software movement, to the dominant view of hackers as "cyberpunk" criminals [3]. A legacy of the early history of hacking in the United States, these associations risk misleading analysts not only about the kinds of individuals who become hackers, but also about the global structure of contemporary hacking [24]. When looking at the diversity of individuals who take part in hacking, one must put aside the image of the hacker as a nerdy teenage boy or girl wearing a black hoody, making computer mischief in the basement of his parents' suburban home. Not only does this stereotype misrepresent the culture of even American hackers, but it also fails to grasp how widely the culture of hackers varies from one country and region to the next.

Conversely, living in a world where access to information is an extremely valuable asset, many will expend extreme measures to either protect or steal this information. The web of networks expanding the globe known as the Internet provides an incredible access to vast amounts of data and information allowing communication as never before. This same interconnectivity, however, also provides a means for others to access information and monitor communications without proper authorization. The architecture of the Internet was driven more by considerations of interoperability and interconnectivity than of information security [14]. Consequently, these systems and infrastructures, and the way people use them, are inherently vulnerable to malicious activity by hackers - individuals who breaks into computers and computer networks to cause harm [59].

This malicious activity can take one of two forms, one (1) destructive in nature [attack] and the other (2) non-destructive through network exploitation which refers to the use of techniques, usually clandestine, to gain unauthorized access, typically to steal information resident on a network [39]. The availability of hacking information and the ease with which illegal actions can be performed allows even amateurs to invade and abuse a network or its users. This dichotomy has been a problem to security professionals and creates opportunities for hackers.

The hacker industry is blurring the distinction between motivated crime and traditional computer security threats - including the disruption of critical infrastructures or the penetration of networks. There have been numerous cyber incidents over the past ranging from attacks against government employees to defense contractors to foreign ministries[1]. A recent study from the Ponemon Institute revealed in its 2021 Cost of Data Breach Report, sponsored by IBM, that the average data breach costs rose from $3.86M in 2020 to $4.24M in 2021, $3.61M was the average cost of a breach in hybrid cloud environments, 38% lost business share of total breach costs, and $4.62M was the average total cost of a ransomware breach[2]. The formalization and commercialization of the hacker industry is producing tools and networks that, while designed to be deployed against corporations or governments, are also capable of targeting individuals [31, 40]. The same techniques and tools that are used to hack into wired systems for purposes of identity theft, fraud or intellectual property theft can be rapidly redeployed to engage in IoT networks as well [26]. In summary, the hacking of IoT networks is enabling an unprecedented form of new cyber-crime.

The IoT landscape of the Internet is dynamic; since IoT devices are generally vulnerable to security threats due to a high-level of transparency, lack of security configurations when implemented, and privacy issues [11]. It wasn't until 1999 that the term "Internet of Things" was first coined by British technology pioneer Kevin Ashton, the Executive Director of Auto-ID Labs at MIT, when he wanted to attract attention to an exciting new technology called radio frequency identification (RFID)[3]. The IoT is changing constantly to accommodate the entry of new smart devices (e.g., tags, readers, sensors, smartphones, etc.), the introduction of fast evolving technologies and methods of access (e.g. Cloud computing, Edge computing, microservices, etc.) and transmissions. Thus, in effect, the IoT is a combination of a technological push and a human pull for more and ever-increasing connectivity with anything happening in the immediate and wider environment – a logical extension of the computing power in a single machine to the environment: the environment as an interface [34].

As the IoT transmission network becomes easier to connect to, exploitation becomes easier to commit due to these devices not being configured for security. Pervasive interconnection has introduced newer and more powerful technologies that may not always meet an IoT provider's reliability and certification standards [60]. These changing markets dynamics have left the door open to an inadvertent or malicious disruption. Interconnection risks are brought on by inexperienced carriers and unproven technologies. Hackers have increased opportunities to exploit IoT network technologies, such as mobile data and Internet protocol-based networks, which have become conduits to valuable transaction traffic and hosted content [33]. As IoT devices increasingly access the Internet, these networks become exposed to new arenas of illicit activities. In turn, as IoT networks become more complex and diversified than those employed today, new protection tools and methodologies will be needed for effective security defense.

## 2. A HISTORY OF HACKING

The Internet has evolved into an open global network which merges all computer networks of the world into one to be accessible for everyone. While the Internet has enabled computer users to share information, it has also brought about negative phenomena, such as computer viruses, pornographic information, illegal access attempts, and theft of confidential information and corruption of internal information [14]. Computer "hackers" today not only possess a full range of attack instruments but have mastered very complicated stealth and evading techniques so sophisticated that they enjoy almost full freedom on the Internet [55].

Hacking has existed, in various forms, since the late 1950s [59]. The term "hacker" has a dual usage in the computer and IT industry today. Researchers originally defined the term as [53]:

> HACKER noun 1). A person who enjoys learning the details of computer systems and how to stretch their capabilities—as opposed to most users of computers, who prefer to learn only the minimum amount necessary. 2). One who programs enthusiastically or who enjoys programming rather than just theorizing about programming.

During the late 1950s, the term 'hacker' was a term created to refer to computer experts in university settings. The original meaning of the word 'hack' started at the Massachusetts Institute of Technology (MIT) and was a term of respect meaning elegant, witty or inspired way making computers work much efficiently [25]. These original 'old school' hackers of the 1950s and 1960s are generally recognized as the ancestors of the modern computer underground [59]. The earliest generations of hackers passionately wanted computers and computers systems designed to be useful and accessible to individuals and, in the process, pioneered public access [38, 42, 61]. Computer programmers from this period saw their work as breaking new ground by challenging old paradigms of computer science [59].

As we entered into the 1960s, computers moved from the university to the military, which angered programmers at the time. To the hackers of the 1960s, secrecy meant the freedom to share code in the computer lab, the spirit of cooperation in program design and the right to tinker with anything and everything based on one's ability to improve upon it [59]. It was during this time that hackers began to create a series of ethics and ideals of hacker ethic's, which defined the tenets as: all information should be free; mistrust authority, promote decentralization; hackers should be judged by their hacking, not by bogus criteria such as degrees, age, race or position; you can create art and beauty on a computer; and, computers can change your life for the better [38].

The 1970s saw the rise of the criminal hacker. Part of the counterculture during this period encouraged phreaking[4] to take money away from capitalist phone companies, the development of the Youth International Party (YIP) to carry out surrealistic subversion and political mischief and the creation of the Technical Assistance Program (TAP) to distribute information freely using the technical aspects of telephony [59]. The first electronic bulletin board systems (BBS) are created during this time to begin to connect the first hacker's on-line [58]. The BBS was a computer system, usually run from a hacker's home, which would function as a kind of community center or public meeting area [58]. At that time, most individuals who had access to computers and modems were programmers or scientists and BBS talk centered on electronics and computers. BBS expanded to a broader range of topics and uses, including the criminal trade of pirated software.

The 1980s solidified the notion of the 'new-school' hacker as a criminal. The personal computer (PC) becomes mainstreamed, increasing the potential hacker population and also increases the number of BBS and modem users [58]. Hacking groups, such as the Legion of Doom/Hackers (USA), Masters of Deception (USA) and the Chaos Computer Club (Germany) develop and become involved in daring and well-known computer break-ins such as online warfare and the jamming of phone lines [61]. The film 'WarGames' demonstrates the potential threat posed by teenagers with computers and introduces the public to hacking [31]. Robert Morris launches the first self-replicating Internet worm (e.g., the computer virus) throughout the nation's computer systems at an alarming rate [59]. Kevin Mitnick[5] and Kevin Lee Pouslen, two hackers who were both arrested, prosecuted and sent to prison for their telephone hacking, gained considered attention in the media about their exploits [59]. The Mentor, an original member of Legion of Doom, writes 'The Hacker Manifesto' which divides the hacker subculture by graphically describing the disposition towards technology that divorces the hacker from the rest of society [59]. These events lead Congress to pass the Computer Fraud and Abuse Acts in 1986[6] and begin to identify the hacking activity as deviant.

With the birth of the Internet, the 1990s saw a dramatic increase in the number of on-line users and potential targets. The film 'Hackers' gave birth to a second generation of hackers and attracted the attention of a new generation of technologically literate hackers who saw the Internet as the next frontier level for exploration [59]. Russian hackers siphon $10 million from Citibank and attack the Pentagon and FBI websites [61]. United States hacker groups are subject to stringent crackdowns by the government [16]. The growth of Wi-Fi and laptops changed the portability of computers and hacker activity [41]. From the beginning of the 21st century until now, the globalization of technology and the Internet expands the link between the hacker communities. Critical infrastructure protection increases in light of growing hacker threats [18]. Sophisticated hacks such as Titan Rain[7] and the TJX[8] demonstrate the increasing threat of the hacker community [64]. Groups such as the Russian Business Network [6] and Shadowcrew [13] demonstrate the reach and professionalism of the hacker industry.

---

[4]Phreaking is using technology or telephone credit cards numbers to avoid long distance charges (Turgeman-Goldschmidt, 2005, p. 32).
[5]The first person convicted under United States law against gaining access to interstate network for criminal purposes (Turgeman-Goldschmidt, 2011, p. 15).
[6]Computer Fraud and Abuse Act of 1986, https://www.congress.gov/bill/99th-congress/house-bi1114718.
[7]Cyber-attacks which originated in China directed against the Defense Information Systems Agency, the Redstone Arsenal, the Army Space and Strategic Defense Installation, and several computer systems critical to military logistics.https://www.everycrsreport.com/reports/RL31787.html.
[8]TJX, the parent company of retailer TJ Maxx, took a $12 million charge in its fiscal first quarter of 2008 due to hacking theft. It cost the company S200 million according to its 2009 SECling. https://www.cio.com/article/272080/intrusion-tjx-takes-12-million-hit-in-first-quarter-for-data-breach.html.

As recently as the early 2000s, most hackers were fundamentally amateurs and were motivated primarily by curiosity, a desire from fame or recognition or a desire to promote a political or social cause (i.e., hacktivism) [47]. These motives, while certainly still present, have been overwhelmed in recent years by the enormous growth of financially motivated cybercrime [43]. In evolving into a lucrative business, hacking has gone from being a niche activity of thrill and glory seeking geeks to becoming an industry that takes a rigorously disciplined approach to innovation and execution [37]. Like all illicit economies, the precise size of the hacking industry is unknown. Although estimates should be taken with a grain of salt, some experts believe that the global market for hacking tools such as bots, adware, spyware and other forms of cybercrime costs the United States economy over $100 billion per year [57]. What is certain is that the value of this marketplace is both growing rapidly and becoming much more sophisticated.

The world has already seen a number of examples of IoT connected devices in the past two decades get hacked [10, 56, 62]. For example, an IBM security intelligence report identified the Jeep hack in July 2015 when a team of researchers were able to take total control of a Jeep SUV using the vehicle's controller area network (CAN bus) [7]. By exploiting a firmware update vulnerability, they hijacked the vehicle over the Sprint cellular network and discovered they could make it speed up, slow down, and even veer off the road [7]. One of the first known IoT attacks against the commercial sector was back in October of 2016; the largest distributed denial-of-service (DDoS) attack ever launched against service provider Dyn using an IoT botnet utilizing malware called Mirai [45]. This led to huge portions of the Internet going down, including Twitter, the Guardian, Netflix, Reddit, and CNN. In early 2016, the FDA confirmed that the implantable cardiac devices made by St. Jude Medical had vulnerabilities that hackers could exploit to access the devices [2]. After gaining control of the device, the hacker could change the pacing of the device, deplete the battery, and even administer shocks to the patient [2]. More recent ransomware IoT attacks have been prolific in targeting supply chains [67], ICS/SCADA systems [67, 69]; and taking down water treatment systems [69], the Colonial Pipeline [70], hospitals [23] and many other industries [1].

A critical element of the IoT hacker industry has been its expanding range of products and services, which are designed both to increase profit margins and to reduce operational risks. As compared with tenyears ago, the IoT hacker industry is not offering a different mix of exploits [39, 57]. Self-propagated worms and hard-drive munching viruses — the dominant concerns of the late 1990s and early 2000s — have given way to intellectual property (IP) and identity theft, spamming, phishing, and denial-of-service (DoS) attacks [8, 15]. This shift has been facilitated by the creation of IoT botnets, which are deployable for a variety of purposes - spamming, DoS, self-replication, cyberwarfare, etc. [70]. Innovation in the IoT hacker industry focuses as much on "social engineering" as on discovering new technical exploits [11]. Hackers are looking for the easiest rather than the cleverest point of entry into systems. Society is still attempting to come to grips with this new criminal activity that knows no geographical boundaries and blurs the notional of criminal jurisdiction [20, 28, 54].

Studies on hacking and hackers have focused on topics such cognitive hacking [19, 27] software piracy [50, 52], hackers' ethics, [17, 32], and application and system programming [4]. Additionally, most studies on hackers have been conducted by practitioners such as researchers, consultants or IT professionals, very often these studies relied heavily on a single interview or proxy populations rather than real hackers [29, 65]. The situation with research related to hacking wireless networks is somewhat more advanced. There are many technical studies related to hacking wireless systems including but not limited to research found in [11, 26, 36]. The research presents empirical evidence of motivations that ethical hackers give for breaching information systems, as well as offering insight into methods for lessening the phenomenon.

Among some of the cited hacking motivations are a perceived technical challenge, a desire for peer recognition, and a desire to educate people about security issues [30, 66].

History has shown that computer networks and its services will have vulnerabilities; all telecommunications, including the IoT, services have security concerns in one fashion or another. Every IoT network has a unique configuration, and its network elements have difference functions, so different security solutions must be implemented. However, it can be argued that IoT networks are more vulnerable to hacking threats and even more so than traditional wireless networks. Although network security is improving with each generation of new technology, it will continue to be an issue for the IoT as more valuable services offered by commerce become available.

## 3. IMPLICATIONS OF THE HACKER PROFESSIONALIZATION

The net effect of the professionalization of the hacker industry is that those interested in tracking hackers cannot afford to pay attention exclusively to the technical innovators (Level 1 [script kiddies] and Level 2 [grey hat] IoT hackers) but must also pay attention to anyone who is technically adept enough to use these much more user-friendly tools (Level 3 [black hat/criminal] IoT hackers, and increasingly, non-technologists). The professionalization of the hacker industry, in other words, has made cyberattacks available to a much wider group of people than ever before. These individuals today are almost exclusively non-state actors, but these hacking platforms can be hired by anyone, and their wide diffusion and ease of use make them potent potential weapons for both state and non-state actors who wish to act either covertly or with plausible deniability.

As the hacker industry becomes more fully professionalized, it is becoming much more adaptive and flexible, making it harder for intelligence and law enforcement to confront. Hacking organizations such as Russia's Strontium (also commonly known as APT28 or Fancy Bear)[9] and China's APT31 (or Zirconium)[10] continue to perform malicious attacks against IoT devices. As with other kinds of illicit industries that have professionalized - drug dealing and human trafficking, for instance - hacking is and will increasingly be dominated by sophisticated actors who understand themselves as 'businessmen', and who regard law enforcement less as a deterrent than as a regulatory force, that is, as a cost of doing business. These managers will reconfigure their businesses quickly in response to intelligence and law enforcement activities. The more law enforcement pressures the hacker industry, the faster it will professionalize and evolve. Therefore, key questions for intelligence and law enforcement include:

- How to balance intelligence-gathering demands (e.g., tracking and monitoring potentially pernicious activities) against a law enforcement demands (e.g., building prosecutions)?

- How to coordinate global law enforcement and intelligence gathering efforts against non-state hackers? In the same way that the United States was able in the early twentieth century to forge an international consensus against narcotics trafficking, it may be possible to build a united international front against hacking.

- How to increase citizen involvement in indicting hacker activity since a distributed response to a distributed problem may in fact be the most effective approach?

---

[9] Microsoft: Russian state hackers are using IoT devices to breach enterprise networks. https://www.zdnet.com/article/microsoft-russian-state-hackers-are-using-iot-devices-to-breach-enterprise-networks/
[10] TheRecord.com. https://therecord.media/chinese-hacking-group-apt31-uses-mesh-of-home-routers-to-disguise-attacks/

The professionalization of the hacker industry, while driven by the increasing predominance of financial motivations, has produced products and services that can easily be redeployed for nonfinancial purposes. For example, while the development of IoT botnets-for-hire (i.e., malware-as-a-service) has been motivated by an effort to expand the market for crimeware and reduce the risk for botnet herders and malware authors, these botnets can be easily directed to disrupting or phishing government systems for ideological or military reasons [11, 68].

## 4. THE EVOLVING HACKER THREAT: HACKING-AS-A-SERVICE

A key concept of this paper is that it is misleading to separate direct threats posed by hacking to United States national security - such as theft of United States government secrets, disruption of critical infrastructure, or penetration of military networks from the rapidly expanding domain of cybercrime against the commercial sector. The same techniques and tools that are used to hack into private systems for purposes of identity theft, fraud, or intellectual property theft can be rapidly redeployed to engage in cyberespionage or cyberwarfare [22]. The rapid formalization, professionalization, and commercialization of the hacker industry is producing tools and networks that, while designed to be deployed against corporations or individuals, are also capable of targeting states and governments[11]. The traditional jurisdictional and analytical distinctions between cybercrime, cyber-espionage, and cyber-sabotage therefore no longer make sense. As security researcher Marcus Willet observed, "Although the SolarWinds hack has been labelled a cyber 'attack', initial analysis indicates that it was intended not to damage, disrupt or destroy networks, but rather to gain intelligence." - a claim that has received great attention with respect to the cyber-security, offensive-cyber and broader national-security policies of the United States and its allies against these types of new form of cyber-attacks[11].

The cost of these new threats from the hacker industry is also enormous. Financial firms flagged nearly $600 million in suspected ransomware payments and United States Treasury investigators have identified billions more[12]. And this estimate does not include the personal or business costs associated with identity theft. To reiterate: the economic costs associated with hacking and the national security threats associated with IoT hacking are best treated as a single unified problem, both from an analytic perspective and from an interdiction perspective. Looking at how different sorts of hacking threats combine not only reveals a bigger threat than commonly conceived, but also provides analytic insight into how this unitary threat is likely to evolve. Although combating cybercrime has traditionally fallen within the jurisdiction of law enforcement, whereas dealing with cyberwarfare and cyberespionage were the responsibility of the intelligence community and the Department of Defense (DoD), the professionalization of the hacker industry has made it important to consider both the criminal and the classic national security dimensions of hacking as part of a single challenge.

A critical element of the professionalization of the hacker industry has been its expanding range of products and services, which are designed both to increase profit margins and to reduce operational risks [11]. As compared with five years ago, the hacker industry is now offering a different mix of exploits; customizing these exploits to local and niche markets; and delivering them not just as products but also as "services" [11]. Innovation cycles within the hacker industry appear to be accelerating within the IoT technology area. Hacker tools are increasingly user friendly and easily downloaded from the Internet (i.e., Offensive Security's Kali LinuxPenetration Tool[13]) lowering the technical barriers to entry for potential hackers. To minimize personal risk, hackers tend not to perform attacks themselves, but instead are simply developing exploits and selling them on to those who will actually use them. Since writing malicious code is rarely illegal (only its malicious application is illegal, not the bits and bytes themselves), those with technical skills are, in effect, moving up the value chain, and away from the legal risks.

---

[11]"Lessons of the SolarWinds Hack", at https://www.iiss.org/blogs/survival-blog/2021/04/lessons-of-the-solarwinds-hack.

[12]"Suspected Ransomware Payments Neraly Doubled This Year, Treasury Says," at https://www.wsj.com/articles/suspected-ransomware-payments-for-first-half-of-2021-total-590-million-11634308503.

[13]The Kali Linux platform contains around 600 penetration-testing programs (tools). https://www.kali.org.

Mirroring the way legal software is increasingly delivered not as a product but as a Web service, hackers are now offering "malware-as-a-service" (MaaS) [21]. Some MaaS sites have become so easy to use customers can use simple Web forms to indicate, for example, which site they wish run a denial-of-service (DoS) attack against, or how many spam emails they wish to send [21]. As security researchers have noted malware authors rent or pass the control of the compromised devices to their peers [9]. Moreover, for many malware families, the attribution of malware to an actor is not straightforward, and due to the malware evolution and code exchange in groups, it becomes a very challenging task. For instance, the MaaSGozi has several variations with different capabilities, with occasional parallel campaigns of its variants to attack banking institutions[14]. The MaaS Emotet, one of the most notorious malwares, is another example of these exchanges designed to steal banking credentials and exfiltrate sensitive information from individual endpoints[15]. In the future, commercialization of crimeware may get to the point where criminals would only get the feed of data from victims that interest them: financial, business, personal, etc., to achieve their malicious objectives.

The development of user-friendly hacking tools that can be operated by non-technical users means that participants in the hacker industry are no longer just Level 1, Level 2, or Level 3 hackers and may in fact not be technologists at all. The participation of non-technologists in the hacker industry has, in turn, strong implications for the evolution of hacker's operations, teaming, and management practices. A critical analytic insight is that the threat posed by hackers is no longer restricted actions executed by technologists, that is, to Level 1, Level 2, and Level 3 hackers. This leaves the actual execution of attacks to others with fewer technical skills but a greater appetite for legal risk. Some of these are as follows:

- As in the legitimate software industry, the technical innovators (Level 1 hackers) are the lodestars of the industry but are not doing the most economically significant work. From an industry perspective, just as important as the Level 1 hackers who create zero-day exploits are the Level 2 hackers who create easy-to-use rootkits, botnets, email lists, etc. - the primary products of the hacker industry.

- As in the legitimate software industry, the hacker industry is not populated only by technologists, but includes people with a variety of specialized skills, who offer a variety of differentiated services - marketing, engineers, salespeople, marketers, tech support, and so on. As often as not, it is non-techies who will use the stolen identities, credit cards, or bank accounts to purchase goods, withdraw money from bank accounts, and launder the money. These downstream activities, crucial to the operations of the hacker industry, entail traditional criminal rather than technical hacking skills.

---

[14]Nngroup. https://research.nccgroup.com/2021/05/04/rm3-curiosities-of-the-wildest-banking-malware/.
[15]Cyberint. https://cyberint.com/blog/research/emotet-returns/.

- As with legitimate software products, many 'business users' of malware may have few if any technical skills. As some researchers note "IoT devices are becoming a favorite target by the attackers due to a lack of security design or implementation. In general, IoT malware has several characteristics such as IoT malware is used to perform DDoS attacks; IoT malware scans the open port of IoT services such as file transfer protocol (FTP), secure shell (SSH) or Telnet; IoT malware performs a brute-force attack to gain access to IoT devices" [51].

## 5. A FUNDAMENTAL UNCERTAINTY

In general, the greater the Internet pipes in a region, the more technologically sophisticated potential hackers can operate. With broadband computers connected full-time to the Internet, the opportunity to attack systems in real time and to use spare bandwidth capacity on compromised machines for further attacks is a resource too tempting to ignore. The reason high levels of bandwidth are necessary is because sophisticated hacking requires broadcasting and receiving enormous amounts of data: for example, using 5G to distribute malware across the world and retrieve the stolen data from infected machines requires a great deal of capacity [12]. High bandwidth, therefore, is usually a necessary requirement for a sophisticated hacker scene to evolve locally, and also makes a tempting target for external hackers. As such, the region in which spam originates may not correspond with the region in which the spammers are located. By contrast, in low-bandwidth regions (e.g., Africa, North Korea, rural Asia, and Latin America) hackers are likely to focus on low-bandwidth intensive exploits, and above all on social engineering attacks [12].

What makes hacking such a unique challenge is precisely this intersection between government/military uses of hacking and the professional, criminal uses of hacking. Although professionalized hackers are primarily profit-driven entrepreneurs, the governments that potentially seek to regulate them are, at bottom, deeply ambivalent about stopping hacking. Although all governments agree in principle that they would like to stop criminal hackers, many governments (including sectors of the United States government) also have a desire to exploit hacking talent and capabilities for offensive military purposes [46].

This fundamental uncertainty makes transnational governmental coordination to interdict hackers and to secure computer systems more challenging than cognate efforts to challenge globalized criminal enterprises, such as the global narcotics industry. While in principle virtually all governments agree that drugs are bad and are usually willing to collaborate even with adversaries to stop production and consumption, it is not at all clear that governments regarding all hacking as bad. They just regard hacking of themselves (and possibly their citizens) as immoral. As with that debate, a decision to "go big" on either cybersecurity or offensive cyber-capabilities not only implies a judgment on the likelihood and effectiveness of alternative threats and modes of war, but fundamentally alters the terms (and perhaps the possibility) of transnational coordination to contain the threat. Given that cyber threats emanate increasingly from non-state actors, compromising the ability of states to coordinate their response is not a decision that should be taken lightly.

What strategy governments use to develop their offensive cyber-capabilities will vary depending on their computing resources? A government like the United States, which has vast resources to devote to development of cyber-capabilities, may be inclined to develop offensive cyberwar capabilities in a centralized manner. Those with fewer resources may choose the "privateering" model. Even without the United States explicitly developing its offensive cyberwar capabilities, a big push on defensive cybersecurity could potentially be regarded as threatening by state adversaries, who would see a much more secure United States as having a strategic advantage. Defensive measures alone, in other words, could themselves provoke a cyber-arms race. Complicating the situation is that potential adversaries include not just states, but also non-state actors, who may not be partners one can negotiate with. Defensive measures designed to counter potential attacks from non-state actors could provoke untoward responses from other states, which might regard these measures as threatening.

## 6. CONCLUSION

What happens as the rest of the world comes online using more IoT devices? IoT Analytics predicts that by 2025, there will likely be more than 27 billion IoT connections"[16]. Given the rapidly lowering barriers to entry for hacking, and the potential lack of sufficient economic opportunities available to the incoming generation of Internet users, the hacker industry should be considered one of the great growth industries of the coming decade, though the demographic of this illicit economy seems likely to change dramatically.

Will organized crime become more centrally involved in hacking? The current level of mafia penetration of the hacker industry is a much-debated topic among security experts and law enforcement regarding this type of behavior [5, 49, 63]. What everyone agrees is that mafia-type organizations are almost certain to become more involved in cybercrime over time[17]. The line between online and offline illicit activities will increasingly blur, as will the distinction between licit and illicit hacking activities. In other words, as organized crime increasingly penetrates the hacker industry, hacking may become merely a single business unit in a diversified criminal organization. The winners, therefore, will be those who discover strategic synergies between their hacking activities and their other activities. The distinction between black hat extortion and white hat security testing services may be increasingly hard to draw.

At the end of the day, the professionalization of hacking means that the United States and other governments face a fundamental tradeoff between law enforcement and military/intelligence appropriation of hacking. Engaging in future-oriented analysis, including scenario-planning, is essential to understand the unintended consequences of variable forms of international interdiction efforts. Some of the risk here is iatrogenic: that is, much of the damage caused by hacking may not come from the hacking itself, but from the efforts of individuals to defend or insulate themselves from hacking. Should attacks like ransomware coming to become a more pervasive problem, we may well witness the balkanization of the Internet, and the withdrawal of Internet users into segmented 'walled gardens'. While this will erode the commercial and political value of the Internet as a universally accessible communications platform, it would alas not end the threat posed by hackers. Hackers would redouble their use of social engineering to continue to infiltrate trust-based online social networks—a process which appears extremely relevant in our society today.

---

[16] IoT Analytics at https://iot-analytics.com/number-connected-iot-devices/.
[17] ScienceDaily.com at https://www.sciencedaily.com/releases/2020/01/200116123805.htm.